\documentstyle[12pt,epsfig,dina4]{article}

\begin{document}
%%wfjm \baselineskip=20pt
%=======================================================================

\begin{center}

{\Large\bf
Breakup Temperature of Target Spectators in
%Temperature for spectator breakup and pre-breakup emission in
\vspace{0.2cm}

$^{197}$Au + $^{197}$Au Collisions at E/A = 1000 MeV}
\vspace*{1.0cm}

\noindent
{
Hongfei~Xi,$^{(1)}$\cite{AAA}
T.~Odeh,$^{(1)}$
R.~Bassini,$^{(2)}$
M.~Begemann-Blaich,$^{(1)}$
A.S.~Botvina,$^{(3)}$\cite{BBB}
S.~Fritz,$^{(1)}$
S.J.~Gaff,$^{(4)}$
C.~Gro\ss,$^{(1)}$
G.~Imm\'{e},$^{(5)}$
I.~Iori,$^{(2)}$
U.~Kleinevo\ss,$^{(1)}$
G.J.~Kunde,$^{(4)}$
W.D.~Kunze,$^{(1)}$
U.~Lynen,$^{(1)}$
V.~Maddalena,$^{(5)}$			%Valentina
M.~Mahi,$^{(1)}$
T.~M\"ohlenkamp,$^{(6)}$
A.~Moroni,$^{(2)}$
W.F.J.~M\"uller,$^{(1)}$
C.~Nociforo,$^{(5)}$       		%Chiara
B.~Ocker,$^{(7)}$
F.~Petruzzelli,$^{(2)}$
J.~Pochodzalla,$^{(8)}$
G.~Raciti,$^{(5)}$
G.~Riccobene,$^{(5)}$			%Giorgio
F.P.~Romano,$^{(5)}$    		%Paolo
Th.~Rubehn,$^{(1)}$
A.~Saija,$^{(5)}$     			%Andrea
M.~Schnittker,$^{(1)}$
A.~Sch\"uttauf,$^{(7)}$
C.~Schwarz,$^{(1)}$
W.~Seidel,$^{(6)}$
V.~Serfling,$^{(1)}$
C.~Sfienti,$^{(5)}$			%Titti
W.~Trautmann,$^{(1)}$
A.~Trzcinski,$^{(9)}$
G.~Verde,$^{(5)}$
A.~W\"orner,$^{(1)}$
and B.~Zwieglinski$^{(9)}$
}

\vspace*{0.5cm}

\noindent
{\it
$^{(1)}$Gesellschaft  f\"ur  Schwerionenforschung, D-64291 Darmstadt, 
Germany\\
$^{(2)}$Istituto di Scienze Fisiche, Universit\`{a} degli Studi
di Milano and I.N.F.N., I-20133 Milano, Italy\\
$^{(3)}$Institute for Nuclear Research,
Russian Academy of Sciences, 117312 Moscow , Russia\\
$^{(4)}$Department of Physics and
Astronomy and National Superconducting Cyclotron Laboratory,
Michigan State University, East Lansing, MI 48824, USA\\
$^{(5)}$Dipartimento di Fisica dell' Universit\`{a}
and I.N.F.N.,
I-95129 Catania, Italy\\
$^{(6)}$Forschungszentrum Rossendorf, D-01314 Dresden, Germany\\
$^{(7)}$Institut f\"ur Kernphysik,
Universit\"at Frankfurt, D-60486 Frankfurt, Germany\\
$^{(8)}$Max-Planck-Institut f\"ur Kernphysik,
D-69117 Heidelberg, Germany\\
$^{(9)}$Soltan Institute for Nuclear Studies,
00-681 Warsaw, Hoza 69, Poland
}

\vspace{0.3cm}

%(draft \today)
\end{center}

\newpage

{\bf
ABSTRACT}
\vspace{0.3cm}

Breakup temperatures were deduced from double ratios of isotope yields 
for target spectators produced in the reaction 
$^{197}$Au + $^{197}$Au at 1000 MeV per nucleon. Pairs of $^{3,4}$He 
and $^{6,7}$Li isotopes and pairs of $^{3,4}$He and H isotopes 
(p, d and d, t) yield consistent temperatures 
after feeding corrections, based on the quantum statistical model, 
are applied. The temperatures rise with decreasing impact parameter from
4 MeV for peripheral to about 10 MeV for the most central collisions.

The good agreement with the breakup temperatures measured previously for
projectile spectators at an incident energy of 600 MeV per
nucleon confirms the universality established for 
the spectator decay at relativistic bombarding 
energies. The measured temperatures also agree with the breakup temperatures
predicted by the statistical multifragmentation model.
For these calculations a relation between the initial 
excitation energy and mass 
was derived which gives good simultaneous agreement 
for the fragment charge correlations.

The energy spectra of light charged particles, measured at
$\theta_{lab}$ = 150$^{\circ}$, exhibit 
Maxwellian shapes with inverse slope parameters much higher than 
the breakup temperatures.
The statistical multifragmentation model, because Coulomb repulsion
and sequential decay 
processes are included, yields light-particle spectra with 
inverse slope parameters higher than the breakup temperatures but 
considerably below the measured values.
The systematic behavior of the differences suggests that they are caused
by light-charged-particle emission 
prior to the final breakup stage.

\vspace{2cm}


\vspace{0.3cm}


\newpage

\pagebreak
\section{Introduction}
\label{Sec_1}

Heavy ion reactions at relativistic energies offer a wide range of
possibilities to study the multi-fragment decay of highly excited 
nuclei [1-8].
%\cite{ogilvie,tsang1,traut,reis97,lips,kwiat,haug96,mowo93}. 
In collisions of heavy nuclei at incident energies exceeding 
values of about 100 MeV per nucleon \cite{tsang1}, 
highly excited and equilibrated spectator systems are formed which 
decay by multifragmentation \cite{schuetti} in good agreement with
statistical predictions \cite{hubel92,botv95}.
The analysis of the kinetic energies
of the decay products has not revealed significant flow 
effects \cite{traut,schuetti,voli}. 
Therefore, the spectator nuclei which are produced over
wide ranges of excitation energy and mass in these reactions, are well suited
for the investigation of highly excited nuclear systems 
in thermodynamical equilibrium.

From the simultaneous measurement of the temperature and the excitation 
energy for excited projectile spectators 
in $^{197}$Au + $^{197}$Au collisions 
at 600 MeV per nucleon, a caloric curve of nuclei has recently been obtained 
\cite{poch95}. For the temperature determination the method suggested
by Albergo {\it et al.} has been
used which is based on the assumption of chemical equilibrium and
requires the measurement of double ratios of 
isotopic yields \cite{albergo}. The obtained
temperatures, plotted against the measured excitation energy, resulted in a
caloric curve with the characteristic behavior reminiscent of 
first-order phase transitions in macroscopic systems.
The 'liquid' and the 'vapor' regimes where
the temperature rises with increasing excitation energy are separated
by a region of nearly constant temperature $T \approx$ 5 MeV over which
the multiplicity of the fragmentation products increases. These results
and, in particular, the apparent rise of the breakup temperature at 
excitation energies exceeding 10 MeV per 
nucleon have initiated a widespread discussion which 
addresses both methodical aspects and questions of interpretation 
[15-22].
%\cite{natowitz,more96,tsang96,campi96,kolomiets1,tsang96b,ma97,wada97}.

The qualitative shape of the caloric curve of nuclei has been
predicted long ago on the basis of the statistical multifragmentation
model \cite{bond85}. The same model, more recently, has been shown to
rather accurately describe the charge correlations measured for the
reaction $^{197}$Au on Cu at $E/A$ = 600 MeV, including
their dispersions around the mean behavior \cite{botv95}. Naturally,
the question arises whether a comparably quantitative level of accuracy
can be reached for the reproduction of the measured caloric curve.
A statistical description of the spectator decay will only be consistent
if the model parameters, including the temperature, are
in the range given by the experiment.
The breakup temperature
should also exhibit the invariance with respect
to the entrance channel that has been
found for the fragmentation patterns.
Their universal dependence on $Z_{bound}$ ($Z_{bound}$ scaling) is a
prominent and well established feature of the spectator decay where 
$Z_{bound}$, defined as the sum of the atomic numbers
$Z_i$ of all projectile fragments with $Z_i \geq$ 2, is a quantity 
closely correlated with the excitation energy imparted to the primary
spectator system \cite{schuetti}.

In this work, we present results of a new measurement of the breakup
temperature in the $^{197}$Au + $^{197}$Au reaction, and
we address these open questions associated
with the statistical interpretation of multi-fragment decays of excited
spectators. High-resolution telescopes were used to measure
isotopically resolved yields of light charged particles and 
intermediate-mass fragments emitted by target spectators produced
at a bombarding energy of 1000 MeV per nucleon.
Methodically, we relied upon the observed isotropy of the spectator 
decay \cite{schuetti} and replaced the measurement of
solid-angle-integrated yields by that of differential yields at selected
angles. Temperatures were deduced from double yield ratios of 
H, He, and Li isotopes whereby feeding corrections, obtained from the quantum 
statistical model \cite{hahn88,konop94}, were applied.

The deduced breakup temperatures are
compared to those measured previously with the ALADIN spectrometer 
for the $^{197}$Au + $^{197}$Au system at
600 MeV per nucleon \cite{poch95,theo} and to 
predictions of a recent version of the
statistical multifragmentation model \cite{bond95}. 
The input parameters for these calculations were derived from the
requirement of a good simultaneous reproduction of
the fragment charge correlations.
While these comparisons turn out to be rather favourable, 
it is evident from the measured energy spectra 
that emission prior to the multi-fragment breakup contributes significantly
to the observed light-particle yields.

\section{Experimental method}
\label{Sec_2}

 The experiment was performed at the ALADIN spectrometer of the GSI
facility \cite{schuetti,hubel91}. Beams of $^{197}$Au of 1000 MeV per
nucleon incident energy were provided by the heavy-ion
synchrotron SIS and directed onto targets of 25-mg/cm$^2$ areal thickness. 
The present data were taken as part of a larger
experiment which incorporated three multi-detector 
hodoscopes for correlated particle detection, 
built at INFN-Catania, GSI, and Michigan State University
and consisting of a total of 216 Si-CsI(Tl) telescopes \cite{poch96,imme}.

A set of seven telescopes, each consisting of three Si detectors with 
thickness 50, 300, and 1000 $\mu$m and of a 4-cm long CsI(Tl) scintillator
with photodiode readout, were used to
measure the isotopic yields of light charged particles and fragments. 
Four telescopes were placed in the forward hemisphere
while three telescopes
were placed at $\theta_{lab}$ = 110$^{\circ}$, 130$^{\circ}$, 
and 150$^{\circ}$ for detecting the products of the target-spectator
decay. Each telescope subtended a solid angle of 7.0 msr.
Permanent magnets were placed next 
to the entrance collimator of each telescope in order to
deflect $\delta$ electrons emerging from the target.

For a global characterization of the reaction and
impact parameter selection, 
the quantity $Z_{bound}$ of the coincident projectile decay was measured
with the time-of-flight (TOF) wall of the ALADIN spectrometer. 
Because of the 
symmetry of the collision system, the mean values of $Z_{bound}$,
for a given event class, should be the
same for the target and the projectile spectators.
Only very small differences may arise from 
the finite dispersion of the relation between $Z_{bound}$ and the
impact parameter \cite{ogil93}.

The energy calibration of the silicon detectors 
was obtained from the calculated punch-through energies for several 
isotopes. Radioactive sources of $\alpha$ particles
were also used. For the CsI(Tl) detectors, energy calibration was achieved
by using the calculated punch-through energies of protons, deuterons, 
and tritons, the calculated energies of particles whose energy loss
was measured with the preceding Si detector, and theoretical light-output
curves \cite{kunde1,milkau1}. 

The quality of the obtained particle identification is illustrated 
in fig. 1 for the case of the telescope placed at 
$\theta_{lab}$ = 150$^{\circ}$.
Isotopes in the range from hydrogen to carbon were
satisfactorily resolved.
Isotopic yields were determined by Gaussian fitting 
with background subtraction. The yields of particles
not stopped in the CsI(Tl) detectors was very low at these backward angles and
did not affect the deduced isotope ratios. 
The contamination of lithium yields by two-$^4$He events 
(decay of $^8$Be) was estimated to be 2\% and ignored.
However, because of the presumably different shapes of
$^3$He and $^4$He spectra at low energies 
(cf. refs. \cite{wada97,boug97,xi97}),
a correction was required in order to compensate for the effect of the
detection threshold $E \approx$ 12 MeV for helium, 
resulting from triggering with the
300-$\mu$m detector. The extrapolation to zero threshold energy
was based on Maxwellian spectra shapes fitted 
to the measured parts of the spectra 
(fig. 2, full lines). The systematic uncertainty
associated with this procedure was assessed by comparing to other
methods of extrapolation. An upper limit of the $^3$He/$^4$He yield ratio
was obtained from a data set composed of events triggered by other 
detectors and extending down to the identification threshold for helium
of $E \approx$ 8 MeV. 
A lower limit was obtained by choosing different energy thresholds 
$E \ge$ 12 MeV 
in the off-line analysis and by linearly extrapolating the resulting isotope
ratios to zero threshold height. The linear extrapolation overestimates
the threshold effect because the particle intensities 
should decrease at very small energies. 
For these extrapolations of the helium yield ratio, the
fitted and background corrected yields as 
obtained from the identification spectra were used. 

For the hydrogen and lithium isotopes, 
extrapolations to zero threshold were not considered necessary 
and the
measured yield ratios above threshold were used. 
The experimental thresholds are sufficiently low for the
hydrogens and, in the lithium case, a significant mass dependence
of the energy spectra was neither expected nor found. 
This is illustrated in fig. 3. The curve derived from fitting the
spectrum of $^{7}$Li ions describes very well also that of $^{6}$Li
(left panel). The $^{6}$Li/$^{7}$Li yield ratio does not change
significantly as a function of the applied threshold energy (right panel).

\section{Experimental results and discussion}
\label{Sec_3}

{\bf 3.1 Experimental temperatures}
\vspace{0.2cm}

Emission temperatures $T$ were deduced from the double ratios $R$ of
yields $Y_i$ measured for pairs of neighboring H, He, or Li isotopes.
Under the assumption of chemical equilibrium, they may be expressed
as
\begin{equation}
R=\frac{Y_{1}/Y_{2}}{Y_{3}/Y_{4}} =
a\cdot \exp(((B_{1}-B_{2})-(B_{3}-B_{4}))/T).
\label{EQ1}
\end{equation}
Here $B_{i}$ denotes the binding energy of particle species i and the 
constant
$a$ contains the ground-state spins and mass numbers.
For the ratios to be sufficiently sensitive to temperature the
double difference of the binding energies
\begin{equation}
b = (B_{1}-B_{2})-(B_{3}-B_{4})
\label{EQ2}
\end{equation}
should be larger than the typical temperature
to be measured \cite{tsang96b}.
In this work we chose $^{3}$He/$^{4}$He as one of the two
ratios (the difference in binding energy is 20.6 MeV)
which was combined with either the lithium yield ratio
$^{6}$Li/$^{7}$Li or with the hydrogen yield ratios p/d or d/t. The
set of $^{3}$He, $^{4}$He, $^{6}$Li, and $^{7}$Li isotopes is the one
used previously for the determination of breakup temperatures of
projectile spectators in $^{197}$Au + $^{197}$Au at 600 MeV
per nucleon \cite{poch95}.
Combinations involving p, d, or t, together with $^{3}$He and $^{4}$He,
have the advantage of larger production cross sections, particularly
in the 'vapor' regime where the
heavier fragments are becoming rare.

The four yield ratios used in this work are shown in fig. 4
as a function of $Z_{bound}$.
For $^{3}$He/$^{4}$He the systematic errors corresponding
to the two limits of extrapolation, described in the previous section,
are indicated by the brackets. The particular sensitivity of this ratio
is evident; with
decreasing $Z_{bound}$, i.e. increasing centrality,
it increases by about one order of magnitude while
the other three ratios exhibit a weak and qualitatively similar
dependence on $Z_{bound}$. 

Solving eq. (1) with
respect to $T$ yields the following three expressions:
\begin{equation}
T_{{\rm HeLi},0}= 13.3/ \ln(2.2\frac{Y_{^{6}{\rm Li}}/Y_{^{7}{\rm Li}}}
{Y_{^{3}{\rm He}}/Y_{^{4}{\rm He}}})
\label{EQ3}
\end{equation}
and
\begin{equation}
T_{{\rm Hepd},0}= 18.4/ \ln(5.6\frac{Y_{^{1}{\rm H}}/Y_{^{2}{\rm H}}}
{Y_{^{3}{\rm He}}/Y_{^{4}{\rm He}}})
\label{EQ4}
\end{equation}
and
\begin{equation}
T_{{\rm Hedt},0}= 14.3/ \ln(1.6\frac{Y_{^{2}{\rm H}}/Y_{^{3}{\rm H}}}
{Y_{^{3}{\rm He}}/Y_{^{4}{\rm He}}})
\label{EQ5}
\end{equation}
where the temperatures are given in units of MeV.
The subscript 0 is meant to indicate that these apparent temperatures,
derived from the measured ground-state populations, may be affected by
feeding of these populations from sequentially decaying 
excited states. 
The consequences of side feeding are presently a subject of 
considerable discussion and are quantitatively investigated 
by several groups with different models [36-39].
%\cite{kolomiets2,xi96,majka,gulm97}.
Here, the required corrections were calculated 
with the quantum statistical model which starts from
chemical equilibrium at a given temperature, density, and neutron-to-proton 
(N/Z) ratio and which includes sequential decay \cite{hahn88,konop94}.
In fig. 5, the three apparent temperatures defined in eqs. (3-5) 
are shown as a function of the 
equilibrium temperature $T_{input}$ for the parameters $N/Z$ = 1.49 
(value of $^{197}$Au) and density
$\rho = 0.3 \cdot \rho_0$ (where $\rho_0$ is the saturation
density of nuclei).  
The relations between $T_{{\rm HeLi},0}$ or $T_{{\rm Hedt},0}$ and 
$T_{input}$ are almost linear and the corrections required
in these two cases are practically identical, except at the highest
temperatures. The linear approximation, indicated by the dotted line,
corresponds to the constant correction factor $T = 1.2\cdot T_{0}$ 
adopted for $T_{{\rm HeLi}}$ in ref. \cite{poch95}. 
The figure also demonstrates that $T_{{\rm Hepd},0}$ is, apparently,
more strongly affected by feeding effects 
at the higher temperatures.
In this case, the double ratio includes the yield of protons which
are likely to be produced in the decay of excited light fragments.
Within the range of densities 0.1 $\le \rho /\rho_0 \le$ 0.5,
the corrections required according to the quantum statistical model 
vary within about $\pm$15\% \cite{theo}.
They virtually do not change with
the $N/Z$ ratio of the primary source.
In the analysis, the corrections calculated for 
$\rho /\rho_0$ = 0.3, as derived from the results 
shown in fig. 5, were applied.

The side-feeding predictions 
obtained by other groups [36-39]
%\cite{kolomiets2,xi96,majka,gulm97}
suggest that the corrections are model dependent. 
In the region of low excitation energies, 
the correction factors required for $T_{{\rm HeLi}}$ range from 1.0 to 1.3, 
within the 
uncertainties of the model assumptions. However, at high excitation 
energies where the consequences of side feeding from
higher lying states may become more important, the results differ 
considerably and depend on the amount of unbound states in the
continuum that are considered \cite{xi96,gulm97} and on the assumed breakup
density \cite{theo,majka,gulm97}.
In the present work, we use the corrections obtained from the quantum 
statistical model in order to be 
consistent with previous analyses. We will also refer
to the statistical multifragmentation model 
which predicts a very similar relation between $T_{{\rm HeLi}}$ 
and the equilibrium temperature (section 3.2).

In fig. 6 the obtained temperatures $T_{{\rm HeLi}}$, 
$T_{{\rm Hepd}}$, and $T_{{\rm Hedt}}$ 
are shown as a function of $Z_{bound}$. They are based on the yield ratios
measured with the telescope at the most backward angle 
$\theta_{lab}$ = 150$^{\circ}$.
Simulations indicate that, at this angle, contributions from the
midrapidity source should be small.
$Z_{bound}$ was determined from the fragments of the projectile decay
measured for the same event.
The temperatures increase continuously with decreasing $Z_{bound}$
from $T$ = 4 MeV for peripheral collisions to about 10 MeV
for the most central collisions associated with the smallest
$Z_{bound}$ values. 
The range $Z_{bound}\le$ 20 corresponds to
the high excitation energies at which the upbend of the temperature appears 
in the caloric curve \cite{poch95}. 
We note here that the cross section for $Z_{bound} <$ 2
is of the order of 130 mb in this reaction \cite{schuetti} so that  
the minimum impact parameter for $Z_{bound} \ge$ 2 (partitions with at 
least one fragment of $Z \ge$ 2) is more than 2 fm in a sharp
cutoff approximation.

The results obtained with the three different double ratios agree rather
well. Only the strong rise of $T_{{\rm HeLi}}$ at small $Z_{bound}$ is
not equally followed by the other two temperatures.
$T_{{\rm HeLi}}$ and $T_{{\rm Hepd}}$ track each other rather closely
which is remarkable in view of the different feeding corrections (fig. 5).
Only $T_{{\rm Hedt}}$ is systematically somewhat 
lower than the other two temperatures. $T_{{\rm Hedt}}$ has also been 
reported to be lower than $T_{{\rm HeLi}}$ in ref. \cite{ma97},
in agreement with the statistical-model calculations of ref. \cite{gulm97}, 
but has been found to be slightly higher than $T_{{\rm HeLi}}$ in
ref. \cite{haug96}.
It seems difficult to assess the precise nature of these 
deviations at the present time. 

A comparison of the $T_{{\rm HeLi}}$ temperatures with
those derived in previous work 
for projectile spectators in $^{197}$Au + $^{197}$Au collisions at
600 MeV per nucleon is given in fig. 7. In the case
of the projectile decay,
the isotopes were identified by tracking of their
trajectories with the upgraded TP-MUSIC detector and subsequent momentum 
and time-of-flight analysis \cite{poch95,theo}.
The displayed data symbols represent the mean values of the range of
systematic uncertainties associated with the two different experiments
while the errors include both statistical
and systematic contributions. The
projectile temperatures are the result of a new analysis of the original
data and are somewhat higher, between 10\% and 20\%,
than those reported previously. Their larger errors follow from a 
reassessment of the potential $^4$He contamination of the $^6$Li yield
caused by $Z$ misidentification.

Within the errors, 
good agreement is observed for the results at 
600 and 1000 MeV per nucleon. 
The expected invariance of the breakup temperature
with the bombarding energy is thus confirmed. 
It is consistent with the $Z_{bound}$ scaling 
of the mean fragment multiplicities and charge correlations 
and supports the statistical interpretation of the multi-fragment 
decay of highly excited spectator nuclei \cite{schuetti}.
The breakup temperatures deduced by the EOS collaboration
for $^{197}$Au + C at 1 GeV per nucleon 
are also consistent with this conclusion \cite{haug96}.
Within the range $Z_{bound} \ge$ 40 which is mainly populated in this
reaction \cite{schuetti}, they are in agreement with the present
results for $^{197}$Au + $^{197}$Au, both in absolute
magnitude and in their dependence on the impact parameter, and thus
confirm the expected invariance with respect to the mass of the 
collision partner.\\

{\bf 3.2 Model calculations}
\vspace{0.2cm}

The calculations within the statistical multifragmentation model
\cite{bond95} were performed in order to test its consistency with
respect to the statistical parameters and predicted charge partitions.
Here one assumes that all observed particles come from the decay of one
equilibrated source.
The correlation of excitation energy and mass of the ensemble of 
excited spectator nuclei, required as input for the calculations,
was chosen in the form shown in fig. 8. The neutron-to-proton 
ratio was taken to be that of $^{197}$Au ($N/Z$ = 1.49).
The distribution of excitation energies
at fixed spectator mass $A_0$ had a Gaussian width,
chosen in proportion to the square root of the mean excitation energy,
and the distribution of masses $A_0$ 
was adjusted in order to reproduce the measured cross section 
$d\sigma/dZ_{bound}$.

As a criterion for setting the excitation energy per nucleon, $E_x/A$, 
we chose the capability of the model to simultaneously
describe the correlations of the mean multiplicity 
$\langle M_{IMF}\rangle$ of intermediate-mass fragments (IMF's) 
and of the mean charge asymmetry $\langle a_{12}\rangle$ with 
$Z_{bound}$. The asymmetry $a_{12}$ of the two largest fragments
is defined as $a_{12} = (Z_{max}-Z_2)/(Z_{max}+Z_2)$,
with the mean value
to be calculated from all events with $Z_{max}\ge Z_2 \ge$ 2.
The comparison, shown in fig. 9, was based on the data reported in 
ref. \cite{schuetti}.
In the region $Z_{bound} >$ 30,
the mean excitation energy of the ensemble of 
spectator nuclei was found to be well
constrained by the mean fragment multiplicity alone.
At $Z_{bound} \approx$ 30 and below, the charge asymmetry 
was a necessary second constraint (cf. ref. \cite{deses96}) while, at
the lowest values of $Z_{bound}$, neither the multiplicity nor
the asymmetry provided rigid constraints on the excitation energy.
These sensitivities are illustrated in fig. 9 where the dashed and
dotted lines show the
model results for $E_x/A$ chosen 15\% above and below the adopted values.

The excitation energies that have resulted from this procedure are
somewhat larger than those found previously in analyses [11, 41-43]
%\cite{botv95,deses96,botv92,barz1} 
of the earlier $^{197}$Au on Cu data at 600 MeV per nucleon
\cite{kreutz}. The difference reflects the sensitivity
to the fragment multiplicity and is caused by the slightly larger
mean multiplicities that were obtained from 
the more recent experiments with improved acceptance \cite{schuetti}.
The excitation energies are still smaller than the experimental
values obtained with the calorimetric method of summing 
up the kinetic energies of the product nuclei
and their mass excess with respect to the ground state of the original
spectator system \cite{schuetti,poch95}.

The calculations proceed such that, for a given mass, charge, 
and excitation energy, first 
the partition function is calculated and then the temperature
$T = \langle T_f \rangle$ is obtained
as the average over the ensemble of partitions.
In the microcanonical approximation, the temperature $T_f$ 
for a particular partition $f$ is found from the energy balance 
\begin{equation}
E_f (T_f,V) = E_0
\label{EQ6}
\end{equation}
where $E_{f}$ is the energy of the system at $T_f$ within the 
volume $V$, and $E_0$ is the total 
available energy \cite{bond95}. At high 
excitation energies, when the system disassembles into many fragments, this
temperature is very close to the grand canonical temperature.

The solid line in fig. 7 represents the thermodynamical
temperature $T$ obtained in this way from the calculations. 
With decreasing $Z_{bound}$, it increases monotonically from
about 5 to 9 MeV. Over a wide range of $Z_{bound}$
it remains close to $T$ = 6 MeV which
reflects the plateau predicted by the statistical multifragmentation model 
for the range of excitation energies 
3 MeV $\le E_x/A \le$ 10 MeV \cite{bond95}.
In model calculations performed for a fixed spectator mass, the plateau
is associated with a strong and monotonic rise of the fragment multiplicities.
Experimentally, due to the decrease of the spectator mass with
increasing excitation energy, the production of intermediate-mass 
fragments passes through a maximum in the corresponding 
range of $Z_{bound}$ of about 20 to 60 (cf. figs. 8, 9).

The dashed line gives the temperature $T_{{\rm HeLi}}$ obtained from 
the calculated isotope yields. Because of sequential feeding, it  
differs from the thermodynamical temperature, the uncorrected 
temperature $T_{{\rm HeLi},0}$ being somewhat lower. 
Here, in order to permit the direct comparison
with the experimental data in one figure, 
we display $T_{{\rm HeLi}}$ which has been 
corrected in the same way with the factor 1.2
suggested by the quantum statistical model. The model $T_{{\rm HeLi}}$
exhibits a more continuous rise with decreasing $Z_{bound}$ 
than the thermodynamical temperature and is in
very good agreement with the measured values.
We thus find that, with the parameters needed to reproduce
the observed charge partitions, this temperature-sensitive observable
is well reproduced. 
In the bin $Z_{bound}\le$ 10 the model values fall below the data. Here,
the experimental uncertainty is rather large but 
the constraint on the excitation energy provided 
by the charge partitions is also rather weak (see above). The discrepancy may
therefore indicate that the excitation energies for breakups corresponding 
to this bin of 
$Z_{bound}$ may be higher than assumed in the calculations.

The correction applied to the calculated $T_{{\rm HeLi}}$ (dashed line) 
results in a good overall agreement with the equilibrium temperature of the 
model calculations (full line). This means that 
the side-feeding corrections suggested by the statistical 
multifragmentation model and by the 
quantum statistical model are qualitatively very similar. 
The difference between them is given by the deviations of the two curves.
They range from about +20\% at large $Z_{bound}$ to -10\% at
small $Z_{bound}$ and thus stay within the range of model uncertainties 
quoted above. They may reflect finite size effects, ignored in the quantum
statistical model, and their variation as the spectator mass changes
as a function of $Z_{bound}$ \cite{poch95}.
They can, equally well, be interpreted as indicating
a variation of the breakup density with $Z_{bound}$. In the region of
small $Z_{bound}$, i.e. at temperatures of about 8 to 10 MeV the correction 
factors of the quantum statistical model are lowered
by 10\% if the density is chosen to be $\rho /\rho_0$ = 0.15 instead of 0.3.
A decrease of the breakup density of that order, as $Z_{bound}$ decreases 
from near 40 to below 20, does not seem unreasonable.
One may therefore expect that an analysis based on measured breakup 
densities (cf. ref. \cite{majka})
will bring $T_{{\rm HeLi}}$ into even better agreement with
the equilibrium temperatures obtained with the statistical
multifragmentation model.\\

{\bf 3.3 Energy spectra}
\vspace{0.2cm}

The kinetic energy spectra were studied for light charged particles up 
to $^{4}$He. 
For the five species proton, deuteron, triton, 
$^{3}$He, and $^{4}$He, the spectra measured at $\theta_{lab}$~=~150$^{\circ}$
and sorted according to 
$Z_{bound}$ are shown in fig. 2. Fit results, 
based on a single target spectator source, are given by the full lines.
They were obtained by assuming that all the particle spectra have 
Maxwellian shapes
\begin{equation}
dN/dE\sim\sqrt{E-V_{c}}\cdot e^{(-(E-V_{C})/T)}
\label{EQ7}
\end{equation}
where $V_{C}$ is the effective Coulomb barrier and $T$ is the temperature.
The smearing of the spectra in the low-energy region due to the finite 
target thickness and a possible motion of the target source (at the most
a few percent of the speed of light, see ref. \cite{schuetti}) were ignored
since their effect on the temperature parameters is too small as to be 
important for the following discussion.

The inverse slope parameters $T_{slope}$ obtained from the fits are shown
in fig. 10 as a function of $Z_{bound}$.  
These kinetic temperatures also increase with decreasing $Z_{bound}$, 
consistent with increasing energy deposition, but their absolute values 
are much higher than the breakup temperatures
deduced from the isotope yield ratios or from the model description.
There is also a tendency to saturate in the $Z_{bound} \le$ 20 region.
Evidently, the inverse slope parameters are
not closely related to the breakup temperatures. 

The slope parameters of the kinetic-energy spectra calculated with the
statistical multifragmentation model are given by the lines shown in 
fig. 10. For the range of $Z_{bound} \le$ 30 they are in the
vicinity of 10 MeV and nearly the same for the five particle species.
This value is considerably higher
than the internal temperatures at breakup
(cf. fig. 7, full line) which reflects the additional 
fluctuations due to Coulomb repulsion and secondary decays after
breakup, to the extent that these effects are incorporated in
the model. Apparently, they account for only part of the 
difference to the experimental slopes.
Towards larger values of $Z_{bound}$, the model results 
start to spread out over the range 5 to 15 MeV, with the protons
exhibiting the lowest and $^3$He the highest
inverse slope parameters.

In fig. 11, the measured kinetic energy spectra, 
integrated over finite ranges of $Z_{bound}$, 
are shown in comparison to the model results.
The two sets of experimental and model spectra are each 
normalized separately, and one overall normalization factor 
is used to relate the two sets.
It was adapted to the yields of $Z$ = 2 fragments
because their calculated multiplicities,
as a function of $Z_{bound}$, are found to satisfactorily 
reproduce the experimental multiplicities reported in ref. \cite{schuetti}.

The main trend apparent from the comparison is a systematically
increasing deviation of the experimental from the model spectra
with decreasing $Z_{bound}$, i.e. increasing centrality, and
with decreasing particle mass. It not only affects the slope parameters
describing the shape of the spectra but also the integrated intensities.
The yields of hydrogen isotopes, and in particular of the protons, are
grossly underestimated by the statistical multifragmentation model.
In the case of $^4$He, on the other hand, 
the equilibrium description accounts rather well for the 
multiplicities and kinetic energies.
A major contribution to the observed $^4$He yields is expected to come
from evaporation by large fragments and excited residue-like 
nuclei which, apparently, is modelled well. 

Conceivable mechanisms that cannot explain the observed deviations
include collective flow and Coulomb effects which both should act
in proportion to the mass or charge of the emitted particle, 
contrary to what is observed. 
On the other hand, the commonly adopted scenario of
freeze-out after expansion involves a pre-breakup phase during which the
system cools not only by adiabatic expansion but also by the emission
of light particles, predominantly nucleons but also light complex
particles \cite{botv92,fried90,papp95}.
The spectra should reflect the
higher temperatures at the earlier stages of the reaction,
prior to the final breakup into fragments.
In addition, there may be contributions from the primary dynamical stage
of the reaction, not as much from the mid-rapidity source itself than from
secondary scatterings with spectator nucleons. 
This picture is in line with the fact that higher
excitation energies were deduced from the experimental data 
\cite{schuetti,poch95} than were 
needed as input for the calculations (fig. 8).
Further investigations of the pre-breakup emission will be needed
in order to clarify this problem (see also refs. \cite{haug96,morl96}). 

A significant component of pre-breakup emission in the light particle
yields has two consequences that deserve particular 
attention. The pre-breakup yields of protons, deuterons, and tritons
are included in the double ratios used to determine the temperatures
$T_{{\rm Hepd}}$ and $T_{{\rm Hedt}}$. This violates the requirement of
thermal and chemical equilibrium, which is the basic assumption
of the method, and thus may shed doubt on the meaning of the consistency 
exhibited in fig. 6. On the other hand, the
deduced temperatures reflect mainly the sensitivity of the $^3$He/$^4$He 
yield ratio. The
p/d ratio varies rather slowly with temperature, as 
evident from fig. 4 and also predicted by the quantum statistical model.
Therefore, the overall p/d ratio and the deduced temperatures
should not be strongly affected by 
contributions to the hydrogen yields from earlier reaction stages.
The second point concerns the excitation energy
carried away prior to the equilibrium breakup. Part of this energy 
may be included in a calorimetric measurement of the spectator
excitation. This has to be taken into account in the interpretation 
of the resulting experimental caloric curve.

\section{Conclusion and outlook}
\label{Sec_4}

Breakup temperatures $T_{{\rm HeLi}}$, $T_{{\rm Hepd}}$, 
and $T_{{\rm Hedt}}$ were measured for target 
spectators in $^{197}$Au + $^{197}$Au collisions at 1000 MeV per nucleon.
In these reactions multifragmentation is the dominant decay channel of 
the produced spectator systems over a wide range of
excitation energy and mass. The corrections for
sequential feeding of the ground-state yields, based on calculations
with the quantum statistical model, resulted in mutually consistent
values for the three temperature observables, except in the range of very
small $Z_{bound}$.

With decreasing $Z_{bound}$, the obtained temperatures increase 
from $T$ = 4 MeV for peripheral collisions to about 10 MeV
for the most central collisions.
Within the errors, the values for $T_{{\rm HeLi}}$
are in good agreement with those measured 
with the ALADIN spectrometer for projectile spectators
in the same reaction at 600 MeV per nucleon. 
This invariance of the breakup temperature
with the bombarding energy is consistent with 
the observed $Z_{bound}$ scaling
of the mean fragment multiplicities and charge correlations 
and supports the statistical interpretation of the multi-fragment 
decay of highly excited spectator nuclei.

The comparison with the results of calculations within the
statistical multifragmentation model shows that a good simultaneous 
agreement for the charge partitions and for the breakup
temperatures can be achieved. A necessary requirement
for a consistent statistical description of the spectator fragmentation
is thus fulfilled.
The obtained equilibrium temperature increases less steeply with 
increasing centrality and stays close to $T$ = 6 MeV 
over a wide range of $Z_{bound}$, coinciding with the
maximum multiplicity of intermediate-mass fragments.
In the bin of smallest $Z_{bound}$, the strong rise of the experimental 
$T_{{\rm HeLi}}$ to 12 MeV is not followed by the model prediction but
neither is the excitation energy, used as input for the model calculations,
well constrained for very small $Z_{bound}$.
The side-feeding corrections obtained with the quantum statistical model
and with the statistical multifragmentation model are in qualitative
agreement. The remaining differences may largely disappear if a
variation of the breakup density with impact parameter is considered
in applying the quantum statistical model. $T_{{\rm HeLi}}$ for small
$Z_{bound}$ will then be lowered accordingly.

The $T_{slope}$ parameters 
characterizing the calculated particle spectra, although higher than
the breakup temperatures, are still considerably smaller than
the experimental values. 
The systematic increase of the deviations with decreasing 
particle mass indicates
that they may be caused by light-particle emission prior to the final
breakup stage. A more quantitative
understanding of the role of the pre-breakup processes will be 
essential for the interpretation of temperatures obtained from 
light-particle yields as well as of the excitation energies
obtained from calorimetric measurements of the spectator source. 
\vspace{0.2cm}

{\it
The authors wish to thank the staff at SIS and GSI for the
excellent working conditions and J. L\"uhning and
W. Quick for technical support. We are grateful to J.~Konopka for providing
us with the results of quantum-statistical-model calculations.
J.P. and M.B. acknowledge the financial support
of the Deutsche Forschungsgemeinschaft under the Contract No. Po 256/2-1
and Be1634/1-1, respectively.
This work was supported by the European Community under
contract ERBFMGECT950083.
}

\newpage

\noindent
{\large\bf
%%wfjm Figure Captions}
Figures}
\vspace{0.5cm}

\centerline{\epsfig{file=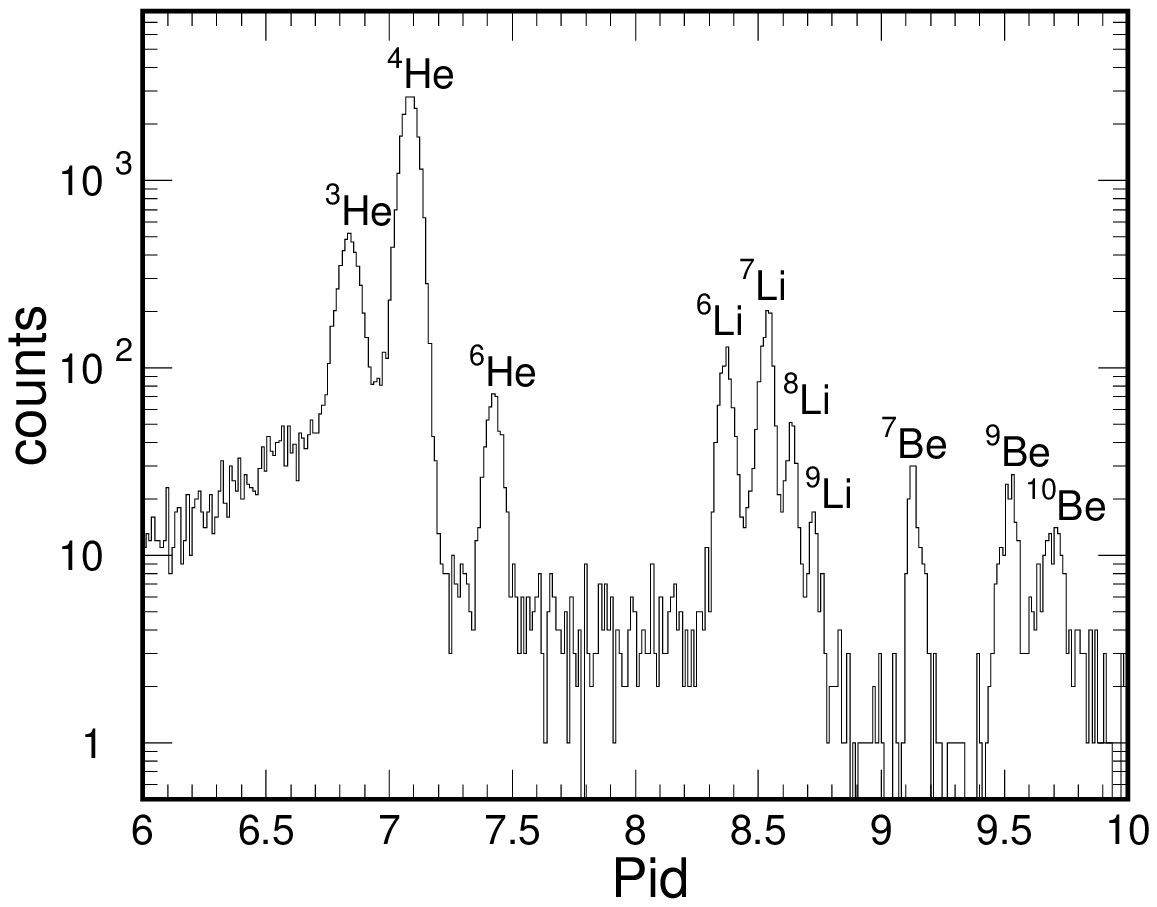,width=\linewidth}}
\noindent
{\bf Fig. 1}.
Spectrum of the particle-identification variable Pid in the range
2 $\le Z \le$ 4 for the telescope
positioned at $\theta_{lab} = 150^{\circ}$. The identified isotopes
are indicated. 
\vspace{0.6cm}

\newpage
\centerline{\epsfig{file=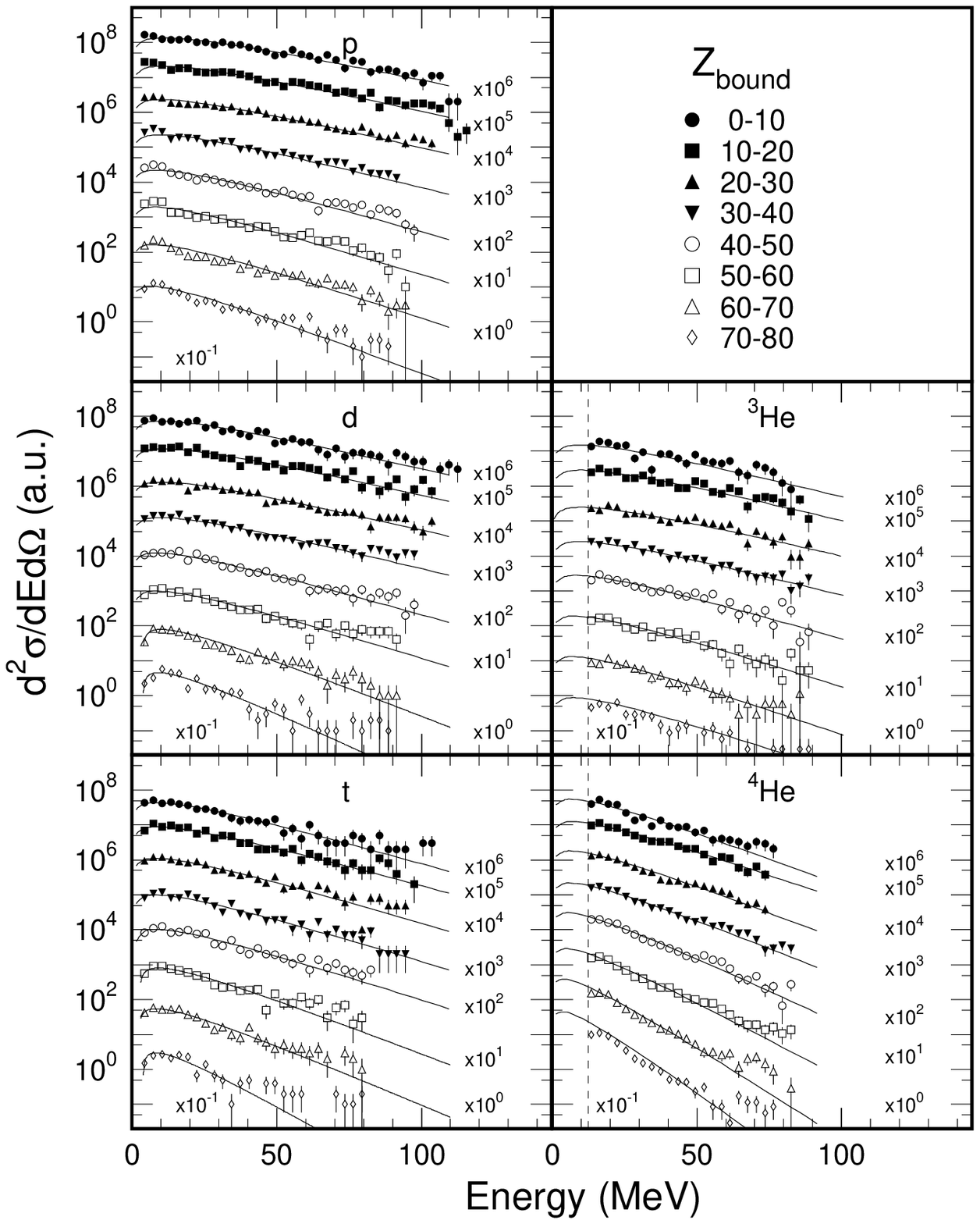,width=0.9\linewidth}}
\noindent
{\bf Fig. 2}.
Energy spectra for hydrogen and helium isotopes, measured at 
$\theta_{lab} = 150^{\circ}$, for 10-unit-wide bins in $Z_{bound}$.
All spectra are 
normalized with respect to each other, note however the scaling factors
of powers of 10. The solid lines are the results of thermal-source fits,
the dashed line indicates the trigger threshold of the telescopes
for helium ions.
\vspace{0.6cm}

\newpage
\centerline{\epsfig{file=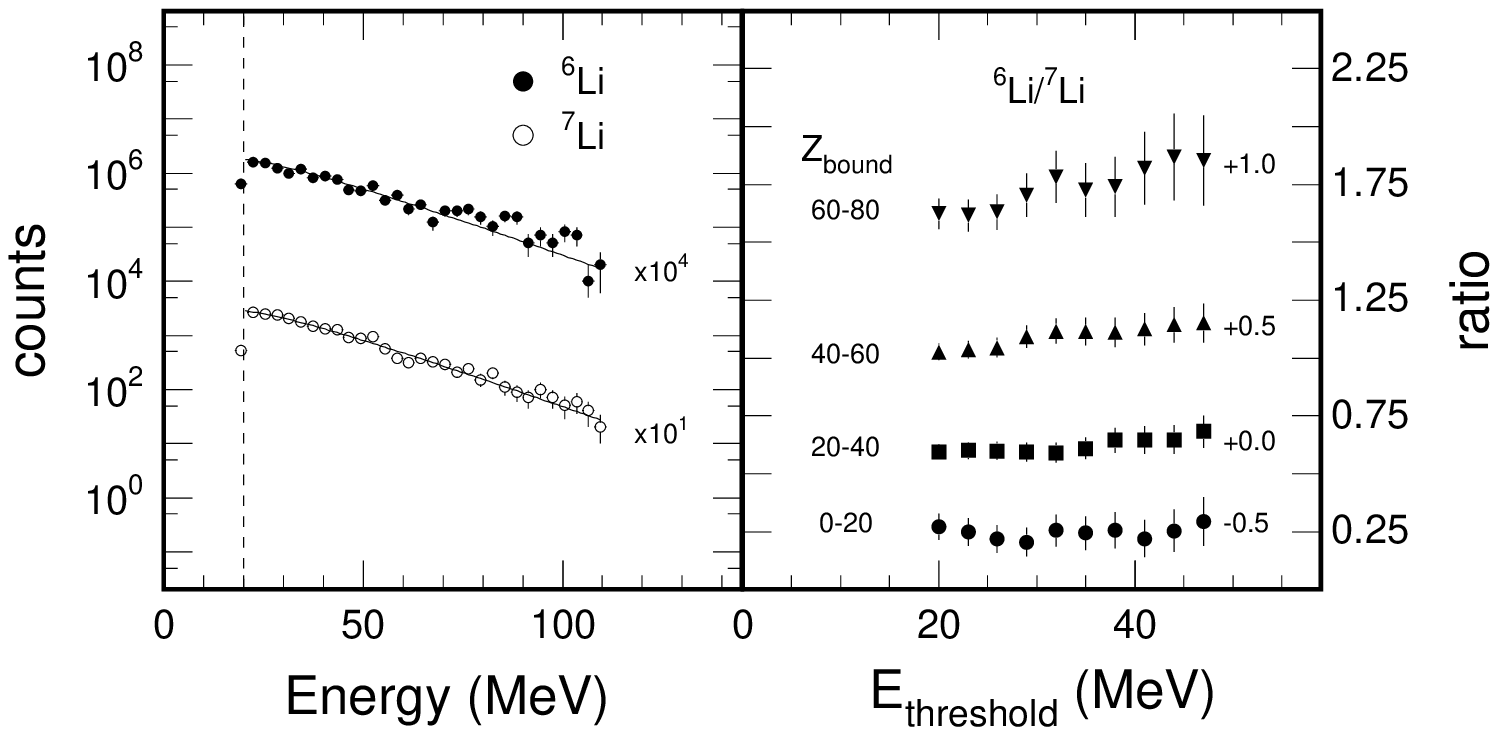,width=\linewidth}}
\noindent
{\bf Fig. 3}.
Energy spectra of $^{6,7}$Li ions, measured at $\theta_{lab} = 150^{\circ}$
(left panel), and their yield ratios 
as a function of the applied energy threshold (right panel).
The energy spectra are integrated over the full range of $Z_{bound}$,
the ratios are given for four bins of $Z_{bound}$ as indicated (note the
linear offsets). The Maxwellian fit of the $^{7}$Li spectrum is 
shown overlaid over the $^{6}$Li spectrum, after an adjustment of 
the overall yield factor (full lines). 
The trigger threshold of the 300-$\mu$m detector
is indicated by the dashed line.
\vspace{0.6cm}

\newpage
\centerline{\epsfig{file=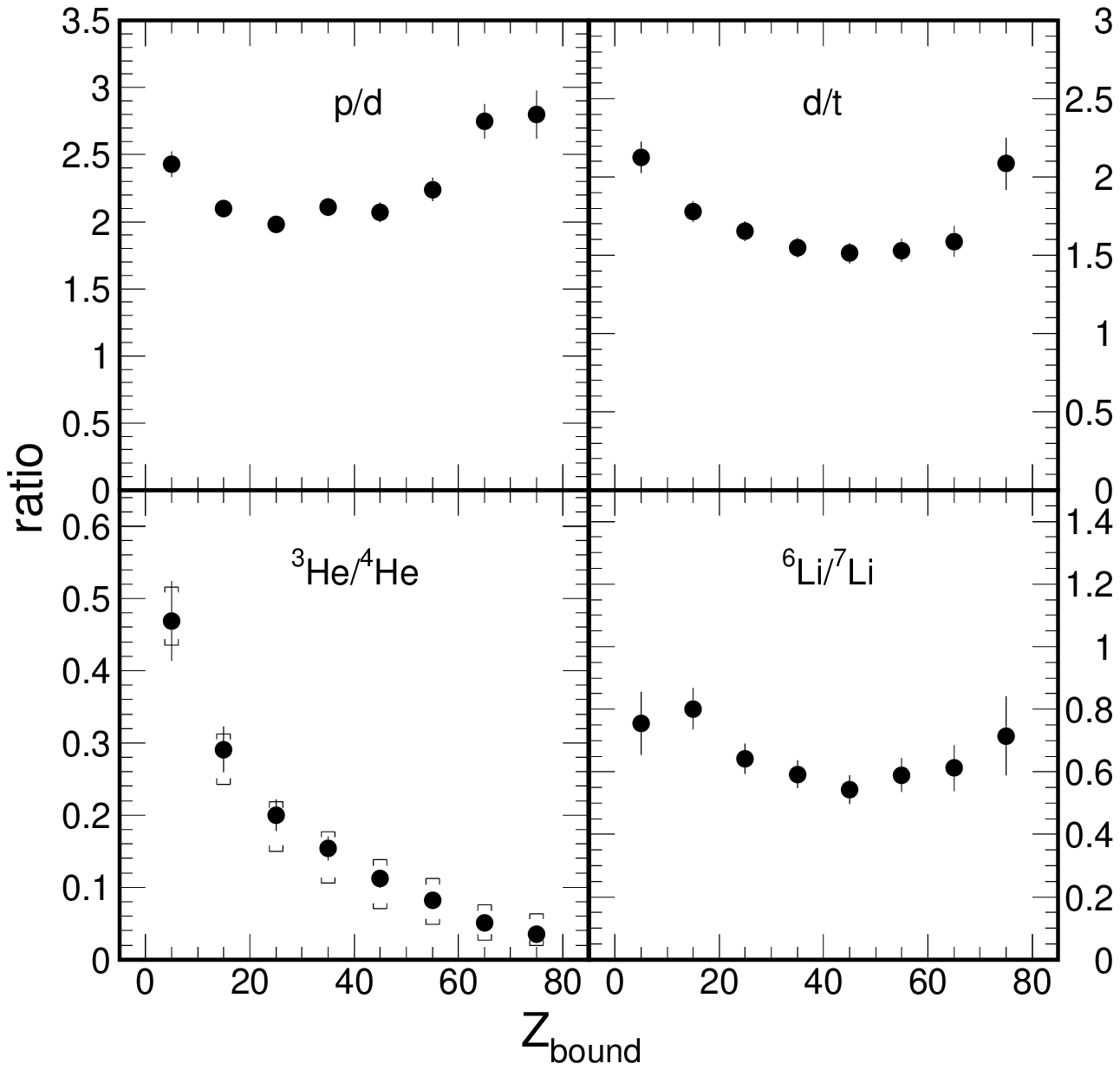,width=\linewidth}}
\noindent
{\bf Fig. 4}.
Four yield ratios of neighbouring H, He, and Li isotopes, measured at 
$\theta_{lab} = 150^{\circ}$ and used to deduce breakup temperatures,
as a function of $Z_{bound}$.
\vspace{0.6cm}

\newpage
\centerline{\epsfig{file=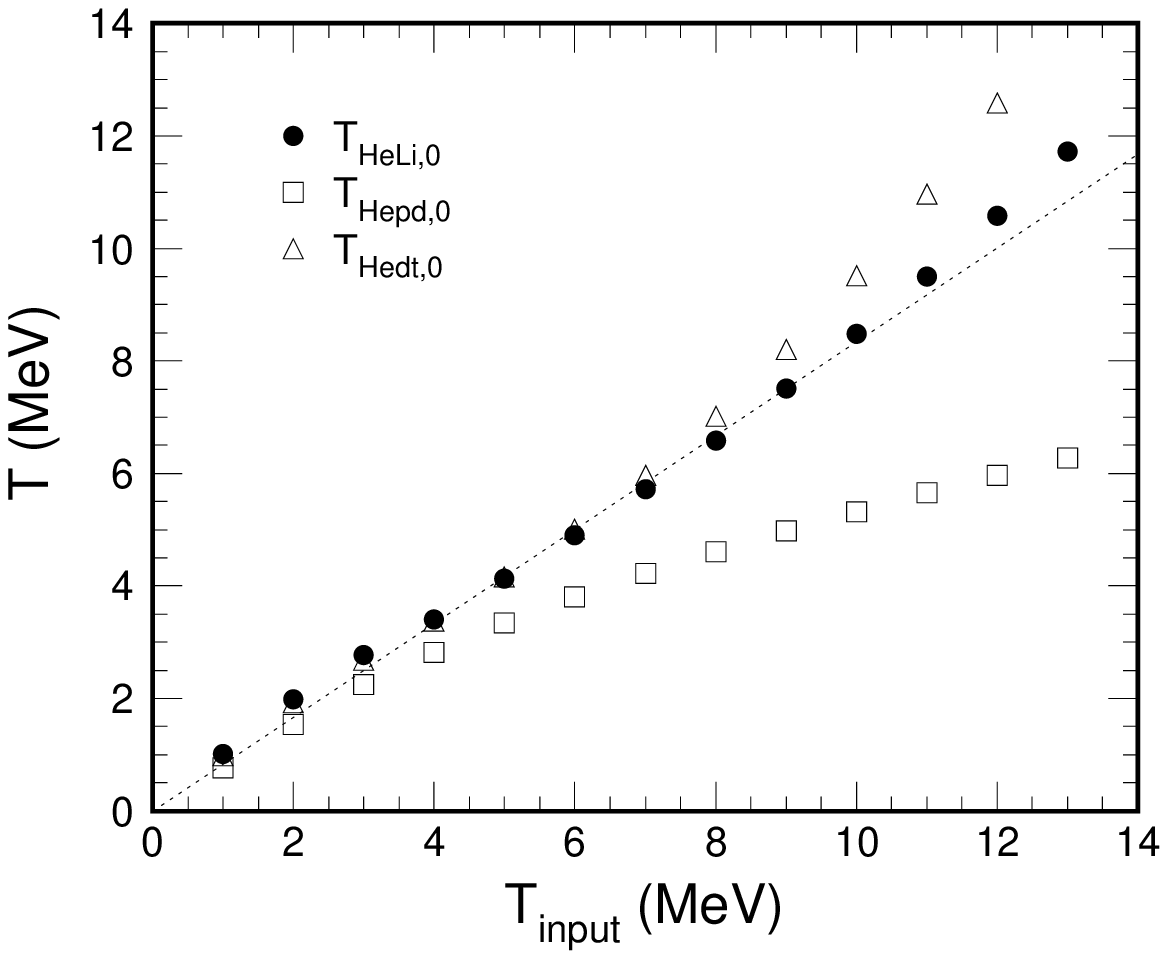,width=\linewidth}}
\noindent
{\bf Fig. 5}.
Apparent temperatures $T_{{\rm HeLi},0}$, $T_{{\rm Hepd},0}$, 
and $T_{{\rm Hedt},0}$, 
according to the quantum statistical model, 
as a function of the input temperature $T_{input}$. A breakup
density $\rho/\rho_0$ = 0.3 is assumed. The dotted line represents the linear
relation $T_0 = T_{input}$/1.2.
\vspace{0.6cm}

\newpage
\centerline{\epsfig{file=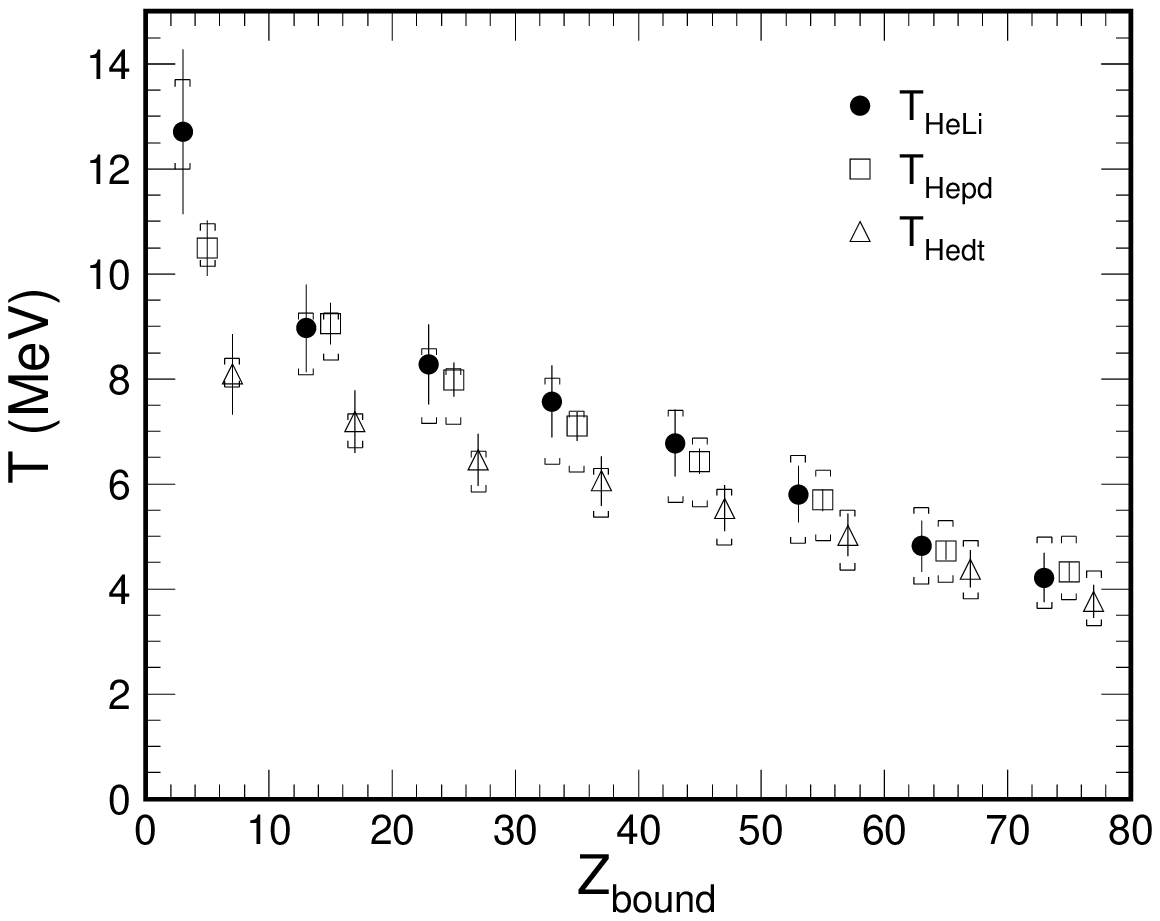,width=\linewidth}}
\noindent
{\bf Fig. 6}.
Temperatures $T_{{\rm HeLi}}$, $T_{{\rm Hepd}}$, and $T_{{\rm Hedt}}$
as a function of $Z_{bound}$, averaged over bins of 10-units width. 
Corrections have been applied as described in the text.
The error bars represent
the statistical uncertainty. The systematic uncertainty, caused by the
extrapolation of the yields of helium isotopes below the identification
threshold, is indicated by the brackets.
For clarity, $T_{{\rm HeLi}}$ and $T_{{\rm Hedt}}$ 
are laterally displaced by 2 units
of $Z_{bound}$.
\vspace{0.6cm}

\newpage
\centerline{\epsfig{file=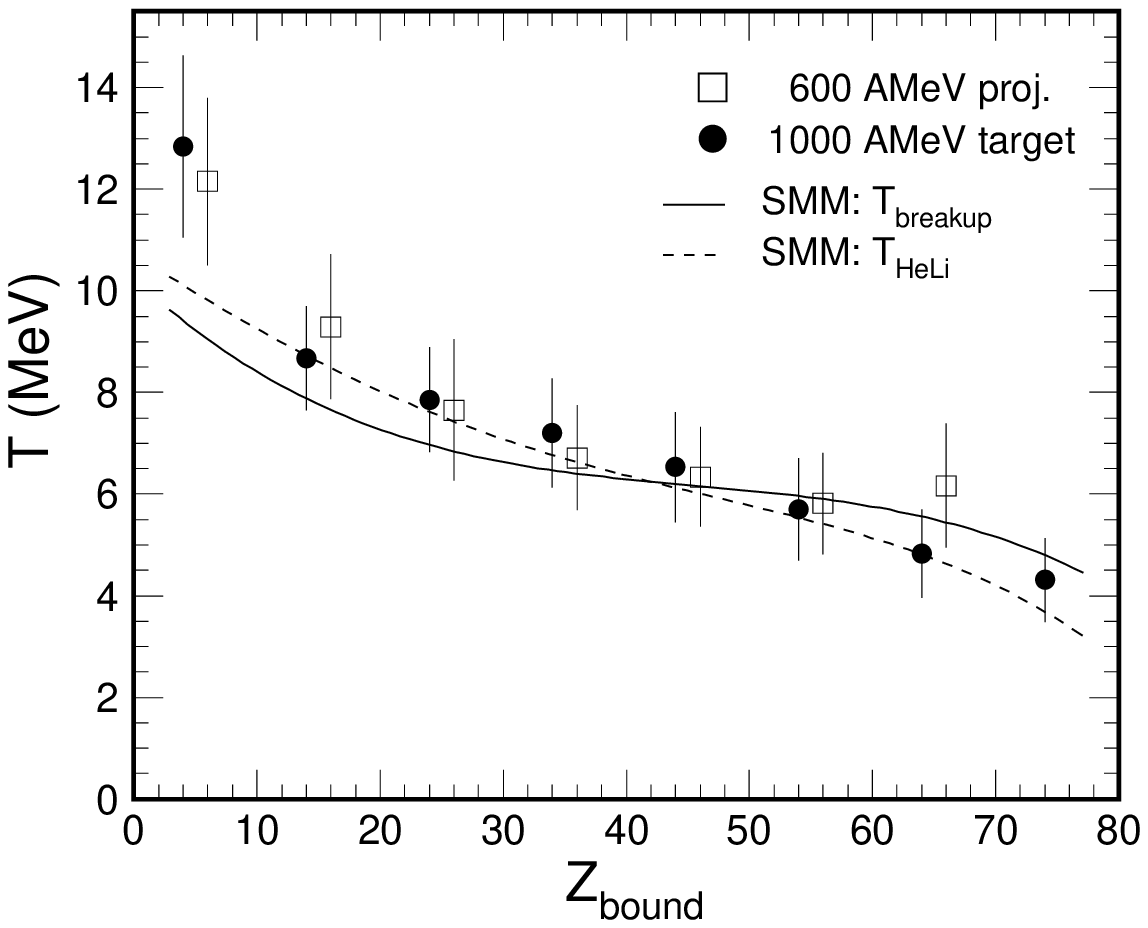,width=\linewidth}}
\noindent
{\bf Fig. 7}.
Temperatures $T_{{\rm HeLi}}$ of the target spectator from the
present experiment at $E/A$ = 1000 MeV (dots)
and of the projectile spectator 
at $E/A$ = 600 MeV (open squares)
as a function of $Z_{bound}$. The data symbols represent averages over
bins of 10-units width and, for clarity, are laterally displaced by 1 unit
of $Z_{bound}$. Statistical and systematic contributions are included in
the displayed errors.
The lines are smoothed fit curves describing 
the breakup temperature $T_{breakup}$ (full line) and the 
isotopic temperature $T_{{\rm HeLi}}$ (dashed line) calculated with the
statistical multifragmentation model.
Note that the trigger threshold affected the data of ref. \cite{schuetti}
at $Z_{bound} \ge$ 65.
\vspace{0.6cm}

\newpage
\centerline{\epsfig{file=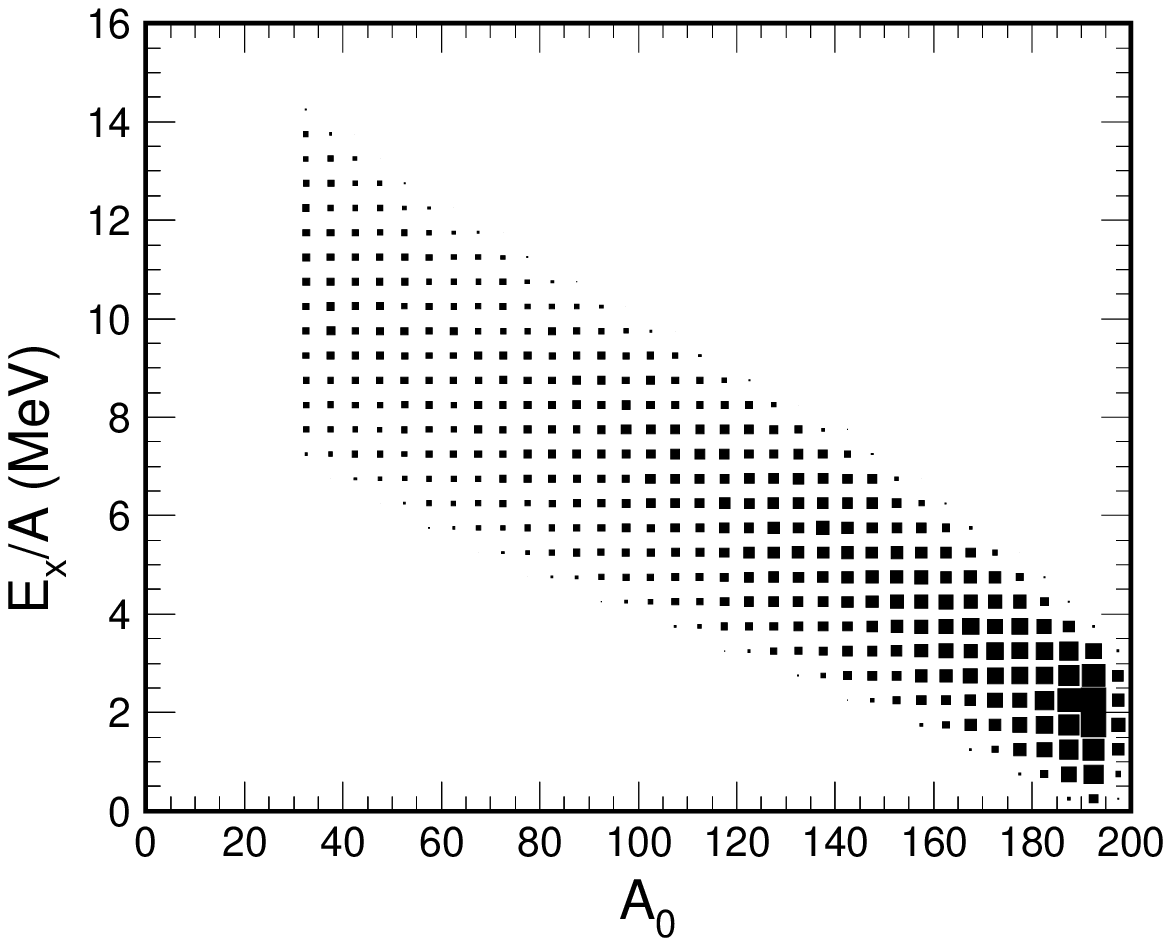,width=\linewidth}}
\noindent
{\bf Fig. 8}.
Excitation energy $E_x/A$ as a function of the mass $A_0$ for the 
ensemble of excited spectator nuclei
used as input for the calculations with the
statistical multifragmentation model. The area of the squares 
is proportional to the intensity.
\vspace{0.6cm}

\newpage
\centerline{\epsfig{file=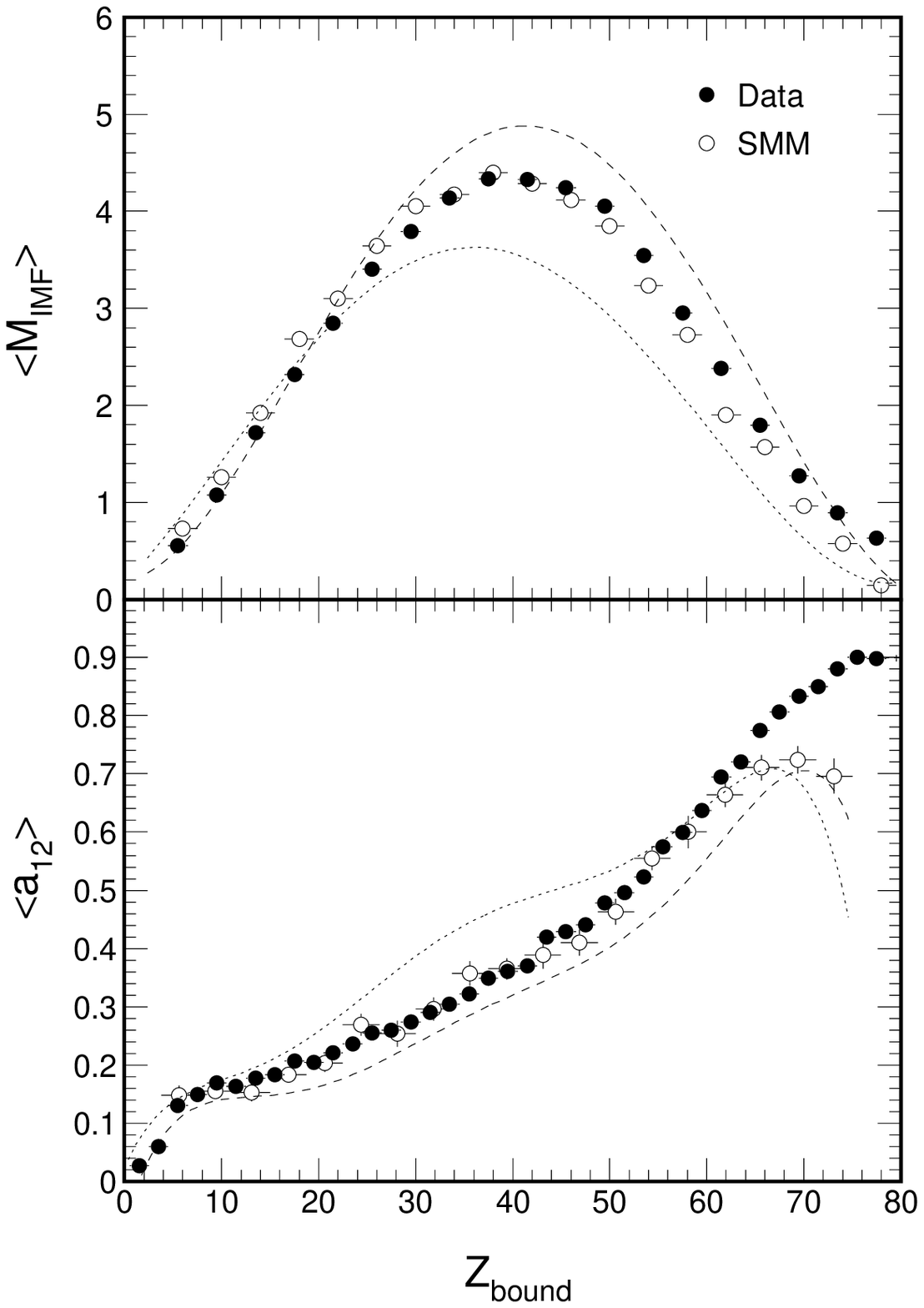,width=0.75\linewidth}}
\noindent
{\bf Fig. 9}.
Mean multiplicity of intermediate-mass
fragments $\langle M_{IMF} \rangle$ (top)
and mean charge asymmetry $\langle a_{12}\rangle$ (bottom) 
as a function of $Z_{bound}$, as obtained from the calculations with the
statistical multifragmentation model (open circles) in comparison to the
experiment (dots, from ref. \cite{schuetti}). The dashed and dotted lines
show the results of the calculations with excitation energies
$E_x/A$ 15\% above and 15\% below the adopted values, respectively.
Note that the trigger threshold affected the data of ref. \cite{schuetti}
at $Z_{bound} \ge$ 65.
\vspace{0.6cm}

\newpage
\centerline{\epsfig{file=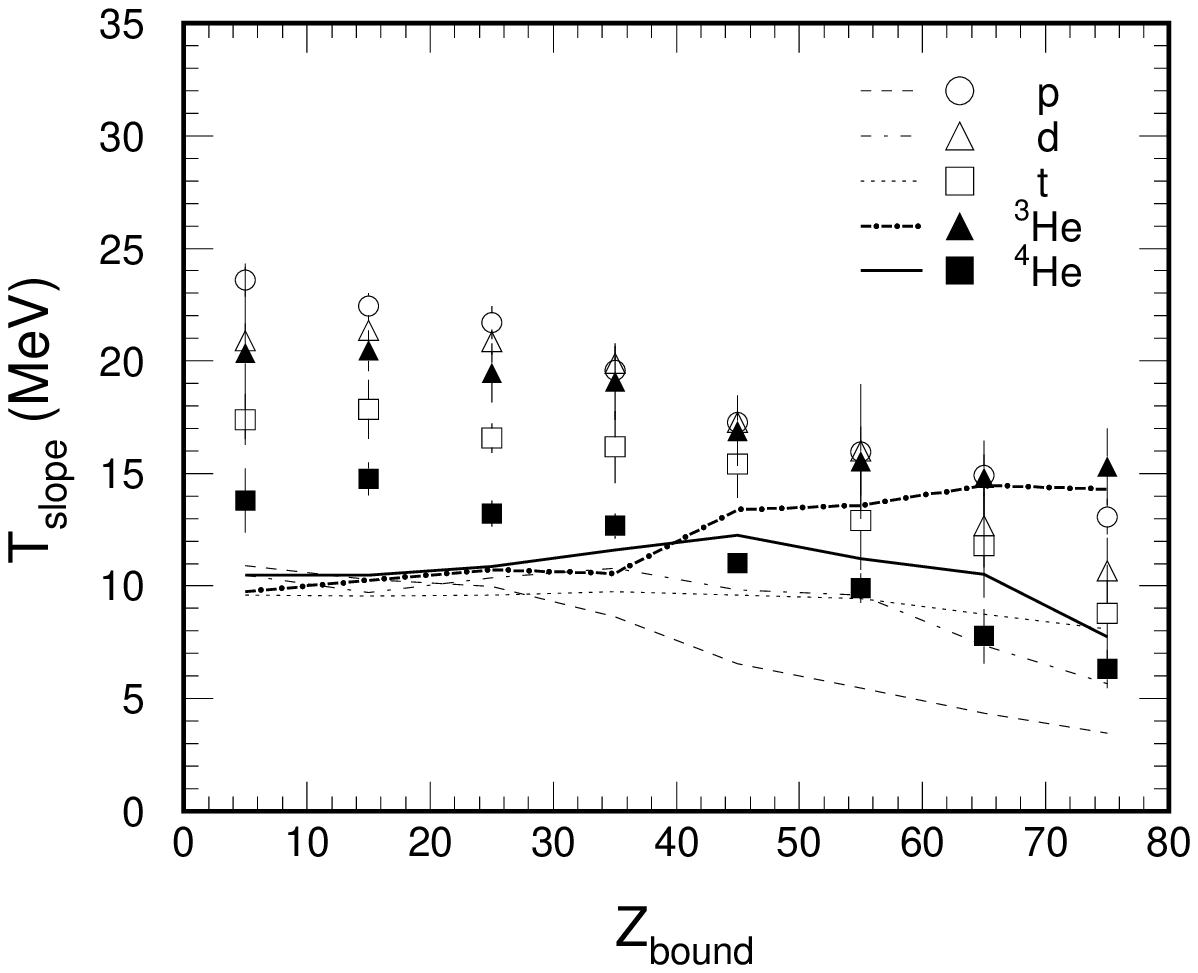,width=\linewidth}}
\noindent
{\bf Fig. 10}.
Inverse slope parameters $T_{slope}$ for hydrogen and helium isotopes,
obtained from fits to the measured (data symbols, 
$\theta_{lab} = 150^{\circ}$) and
calculated (lines) energy spectra, as a function of $Z_{bound}$.
\vspace{0.6cm}

\newpage
\centerline{\epsfig{file=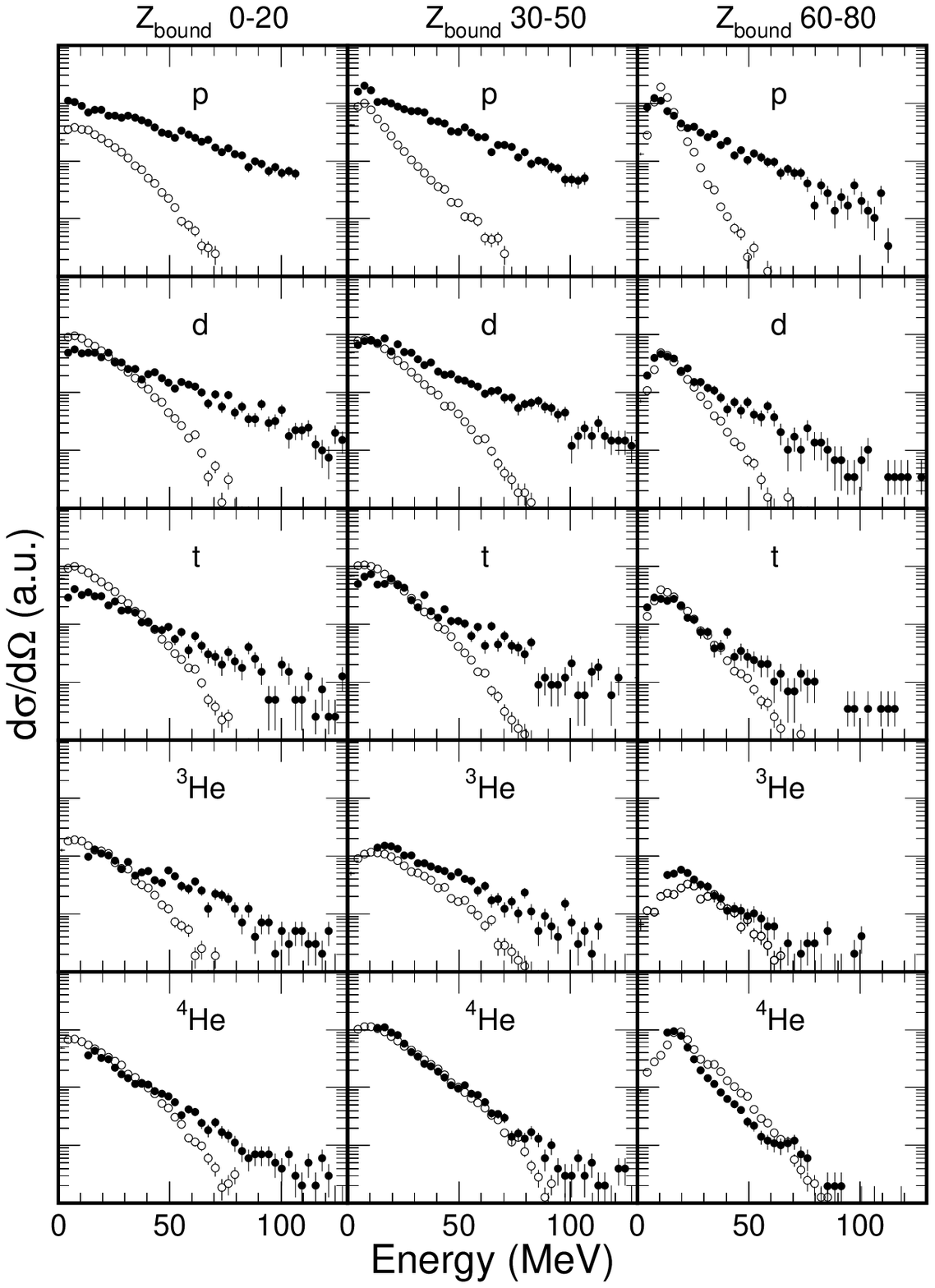,width=0.8\linewidth}}
\noindent
{\bf Fig. 11}.
Energy spectra, measured at $\theta_{lab} = 150^{\circ}$,
of light charged particles p, d, t, $^3$He, 
and $^4$He for three
intervals of $Z_{bound}$ as indicated. The dots represent the
measured spectra, the open circles are the results of the calculations
with the statistical multifragmentation model. The spectra are
normalized as stated in the text.

\end{document}